\documentstyle[12pt,epsf]{article}
\begin{document}
\tolerance=10000
\title{Phase diagram of the 2D $^4$He in the density-temperature plane}
\author{F.~V. Kusmartsev$^a$ and M. Saarela$^b$;
$^a$Nordita, Blegdamsvej 17, Copenhagen, Denmark,
$^b$Department of Physical Sciences,
 Theoretical Physics,
University of Oulu, SF-90570 Oulu, Finland}
\maketitle

\begin{abstract}
Thin $^4$He films adsorbed to weakly attractive substrates form nearly
2D layers. We describe the vortices in 2D superfluid $^4$He
 like quasiparticles. With the aid of   a variational many-body
calculation we estimate their inertial mass and describe their
 interactions with the $^4$He
particles and other vortices. Third sound measurements  revealed
anomalous behavior below the BKT-phase transition temperature. We
ascribe  this to the sound mode traveling in the fluid of
vortex-antivortex pairs. These pairs forms a crystal (or liquid crystal)
 when the film
thickness increases, the third sound mode splits into two
separate modes as seen in  experiments. Our many-body calculation
predicts the critical density, at which 
the phase transition into the vortex-antivortex
state at zero temperature occurs. We also describe the phase diagram of thin
$^4$He films. 
\end{abstract}

\vskip 0.5 truecm

    Contributed paper to the  XXI International Conference on Low Temperature
Physics, August 8-14, 1996.

\vfill
\eject

In thin $^4$He films the superfluidity is distroyed by the creation of
free vortices at finite temperature. Vortices appear from the
dissociation of bound vortex-antivortex (V-A) pairs \cite{BKT} in the
Beresinski-Kosterlitz-Thouless (BKT) phase transition.
The interaction between classical vortex  and antivortex  is attractive
 and logarithmic leading to annihilation at zero separation distance.
The (V-A) pair may exist in motion when the logarithmic attraction
 is balanced by a repulsion due to the Magnus force.

Recently a new transverse third sound mode in the superfluid $^4$He
film was observed \cite{Moch} suggesting the
formation of a crystal phase in the region near
the BKT phase transition. The authors
removed the possibility of substrate corrugation effects and state that
the new phase is observable from zero temperature upto the
BKT-transition temperature at densities below 0.6$\pm$0.1 layers of
$^4$He adsorbed on the solid hydrogen substrate. 

Zhang \cite{Zhan} considered the thermodynamics of vortices and
antivortices by mapping them into a classical two-dimensional (2D)
Coulomb gas and proposed a phase diagram 
with two new phases; the square and hexatic V-A lattice.
 Wagner and Ceperley
\cite{WagnerLT93} made a path integral Monte Carlo study of the film
and found no new crystal  phases in their simulations. 

We have proposed a quantum mechanical approach for the description of
 vortices  as quasiparticles having an inertial mass \cite{SaKu}. In
that variational many-body formalism 
vortices and antivortices are included in the superfluid $^4$He
through the phase factor of the many-body wave function,
\begin{eqnarray}
        \Psi({\bf r}_v,{\bf r}_a,{\bf r}_1...,{\bf r}_N)
         &=& e^{i\sum_{j=1}^N\phi({\bf r}_v,{\bf r}_j)
	- \phi({\bf r}_a,{\bf r}_j)}
\nonumber\\
        &\times&\psi_0({\bf r}_v,{\bf r}_a,{\bf r}_1,...,{\bf r}_N).
\label{wavev}
\end{eqnarray}
The coordinates ${\bf r}_1,...,{\bf r}_N$ refer 
to the superfluid particles and ${\bf r}_v$ and ${\bf r}_a$ to the vortex
and antivortex cores respectively.
The modulus, $\psi_0$, is expanded in terms of the
Jastrow type correlation functions.

The quasiparticle nature of the vortex implies two
assumptions: 
(1) Vortices carry the {\it inertial mass},
which we set equal to the mass
of the expelled superfluid and calculate 
self-consistently 
from the long wave length limit of the vortex-background
structure function. 
(2) The effective interaction between a vortex and background particles
and vortices themselves is determined by the phase factor of the many
body wave
functions plus  the induced interaction caused by the polarization of the
medium due to the presence of the vortices.

These assumption define the effective Hamil\-to\-nian. We calculate the
variational upper bound of its expectation value i.e. the  energy
required to create one V-A pair at zero temperature by
minimizing the vortex-background and vortex-antivortex correlation
functions in  $\psi_0$. The locations of the vortex and antivortex cores
are translationally invariant, and thus the average density of the
superfluid remains constant. The solutions of the optimizing Euler
equations give then  the probability of finding a $^4$He particle at a
given distance away from the vortex core and 
the probability distribution
of the V-A pair at the zero temperature.
\begin{table}
\caption{
The vortex  mass 
and the binding energy of the vortex-antivortex pair,
$\mu^{va}$, as a function of density.
}
\begin{tabular}{cccc}
\hline
& density \AA$^{-2}$
& mass (amu)
& $\mu^{va} (K) $
\\
\hline
&0.035	&  16.67  &-0.57 \\
&0.040	&  8.52   & 0.73 \\
&0.045	&  5.27   & 2.21 \\
&0.050	&  3.59   & 3.84 \\
&0.055	&  2.60   & 5.66 \\
&0.060	&  1.96   & 7.46 \\
&0.065	&  1.53   & 9.39 \\
\hline
\end{tabular}
\label{tab}
\end{table}

As a consequence the distance between the vortex and antivortex is 
determined by the balance between their kinetic energies and
effective interaction giving rise to a bound quantum
vortex-antivortex state. For an illustration the V-A pairs can be
thought as an equivalent to excitons in solids. The breaking up of 
V-A pairs in 2D  $^4$He is similar to the decay of
excitons in Coulomb plasma into  positive and negative charges (
i.e. electron-hole liquid), when the density $\rho_{\rm VA}$ increases. 
When the radius of the Debye screening $\sim 1/\rho_{\rm VA}$ decreases,
 the vortex-antivortex interaction decreases, too, letting
single vortices move independently.

In the careful analysis of the structure of thin
films \cite{filmstruc} one can show that below the density
0.031\AA$^{-2}$ the film becomes unstable against cluster formation,
which is the spinodal instability. Above the density 0.068\AA$^{-2}$ 2D
$^4$He crystalizes, but the solid hydrogen substrate
potential is so weak that atoms jump to a new layer before the first
layer solidifies. This new layer again consists of clusters before the
density is high enough to form a uniform superfluid layer and the
procedure is repeated for thicker films. 

In Table I we give results for the liquid densities of 2D
$^4$He using the Aziz potential between $^4$He atoms and setting the
vorticity equal to one. The V-A pair is assumed to be in
the relative s-state. We find that at densities less than 0.037
\AA$^{-2}$ the chemical potential for creation of 
V-A pair becomes negative. This signals a new kind of an
instability in the system, where the quantum pairs are spontaneously
created.

We argue that by increasing the
fugacity of the pair $y_{\rm VA}$ ( but not of a single vortex
as argued by Zhang \cite{Zhan}) at a fixed  low
temperature (which can be done by decreasing the $^4$He density), the 
density of V-A pairs increases and then it is preferable for the 
pairs to form a lattice before they break due to Debye screening. 
Because of the anisotropy of the V-A excitons
a liquid crystal may be created as well. Our estimate of the critical
density at T = 0, $\rho_c \sim$ 0.5 layers agrees well with
experiments\cite {Moch}. Thus our approach explains the experiments in
the whole temperature as well as the coverage range and gives the
phase diagram shown in Fig. 1.
The phase diagram of a thin $^4$He film.

\ \\

The work was supported by the Academy of Finland (to M.S.), and
NORDITA and Phys. Dep., Loughboro' University, Loughboro', Leicestershire, LE11 3TU, UK (to F.V.K.)
\vfill
\eject
%
%

\end{document}